\documentclass[%
 aip,
cp,  
 amsmath,amssymb,
 reprint,%
]{revtex4-2}

\usepackage{graphicx}
\usepackage{dcolumn}
\usepackage{bm}

\usepackage[utf8]{inputenc}
\usepackage[T1]{fontenc}
\usepackage{mathptmx} 

\begin{document}

\title{Euclidean Path Integral, Entanglement Entropy, and Quantum Boundary Conditions}

\author{Dong-han Yeom} 
 \email[Corresponding author: ]{innocent.yeom@gmail.com}
\affiliation{
Department of Physics Education, Pusan National University, Busan 46241, Republic of Korea\\
Research Center for Dielectric and Advanced Matter Physics, Pusan National University, Busan 46241, Republic of Korea
}

\date{\today} 

\begin{abstract}
To understand the information loss paradox in a consistent way, we provide a brief big picture that describes both outside and inside a black hole. We summary several ideas including the Euclidean path integral, the entanglement entropy, and the quantum gravitational treatment for the singularity. This integrated discussion can provide an alternative point of view toward the ultimate resolution of the information loss paradox.\footnote{Proceedings of the 17th Italian-Korean Symposium on Relativistic Astrophysics. Talk on August 4, 2021, Gunsan, Korea.}
\end{abstract}

\maketitle

\section{Introduction: key ideas to obtain the Page curve}

The information loss paradox of black holes is an unresolved problem in modern theoretical physics \cite{Hawking:1976ra}. There have been fifty years history of the debating \cite{Chen:2014jwq}, but recently there was a big improvement in string theory \cite{Almheiri:2019psf}. In the recent discussion, the basic idea was originated from the computation of the entanglement entropy. The entanglement entropy can be computed from the extreme of the generalized entropy. As a result, we obtain several saddles that are known to be extremal surfaces. If there exist more than one extremal surface, the contribution from the new saddle can explain the Page curve that is deeply related to the unitary evolution of the black hole. The contribution from the new saddle can be justified by the replica trick; the new contribution comes from the replica wormhole \cite{Almheiri:2019qdq}.

However, it is still premature to conclude that we have a sufficient justification of new saddles. A new contribution from the replica trick comes from the path integral of the replicas of the density matrix rather than quantum states. However, this is not a path integral of the quantum state which is more orthodox in quantum mechanics. Therefore, the physical interpretation is vague; how can information be emitted from the black hole after the Page time?

In this regards, to understand the physics in the common ground, we assume the following two contents \cite{Chen:2021jzx}.
\begin{itemize}
\item[--] 1. \textit{Multi-history condition}: there exist at least two solutions (saddles, steepest-descents, or whatever) that dominantly contribute to the entanglement entropy computation, say $h_{1}$ and $h_{2}$. We further assume that $h_{1}$ is an information-losing solution while $h_{2}$ is an information-preserving solution.
\item[--] 2. \textit{Late-time dominance condition}: initially, the probability of $h_{1}$, say $p_{1}$, was dominated. However, as time goes on, the probability of $h_{2}$, say $p_{2}$, is dominated. (The time can be measured in terms of the entropy, or any methods.)
\end{itemize}

Based on these assumptions, the expectation value of the entanglement entropy is approximately \cite{Chen:2021jzx}
\begin{eqnarray}
\langle S \rangle \simeq p_{1} S_{1} + p_{2} S_{2},
\end{eqnarray}
where $S_{1,2}$ denote the entanglement entropy for $h_{1,2}$, respectively. Initially $p_{2} \ll p_{1} \simeq 1$ and eventually $p_{1} \ll p_{2} \simeq 1$. $S_{1}$ monotonically increases or never approaches to zero, while $S_{2}$ eventually decreases to zero. Therefore, one can explain the Page curve for a unitary evolution.

Perhaps, we may interpret as follows. If the recent development is the correct answer to the information loss paradox, then there must exist the state-level description. In other words, in the path-integral of quantum states, there must be a corresponding steepest-descent of the replica wormhole contributions; if it is not possible, then at least, there must be an analog solution that mimics the role of the replica wormhole. Therefore, it might be interesting to understand the entire wave function by using the Euclidean path integral approach \cite{Hartle:1983ai}, i.e., in terms of the state-level path integral.

\section{Instantons for Hawking radiation and\\ tunneling to trivial geometry}

In the Euclidean path integral approach \cite{Hartle:1983ai}, from the past infinity $(h^{\mathrm{in}}_{ab}, \phi^{\mathrm{in}})$ to the future infinity $(h^{\mathrm{out}}_{ab}, \phi^{\mathrm{out}})$, one can provide the propagator by using the following path-integral
\begin{eqnarray}
\Psi_{0} \left[ h^{\mathrm{out}}_{ab}, \phi^{\mathrm{out}}; h^{\mathrm{in}}_{ab}, \phi^{\mathrm{in}} \right] = \int \mathcal{D}g_{\mu\nu} \mathcal{D}\phi \;\; e^{- S_{\mathrm{E}}[g_{\mu\nu},\phi]},
\end{eqnarray}
where we sum-over all $g_{\mu\nu}$ and $\phi$ that connects from $(h^{\mathrm{in}}_{ab}, \phi^{\mathrm{in}})$ to $(h^{\mathrm{out}}_{ab}, \phi^{\mathrm{out}})$. This Euclidean path-integral can be approximated by the steepest-descent approximation:
\begin{eqnarray}
\Psi_{0} \left[ h^{\mathrm{out}}_{ab}, \phi^{\mathrm{out}}; h^{\mathrm{in}}_{ab}, \phi^{\mathrm{in}} \right] \simeq \sum_{\mathrm{on-shell}} e^{- S_{\mathrm{E}}^{\mathrm{on-shell}}[g_{\mu\nu},\phi]},
\end{eqnarray}
where we sum-over all on-shell solutions, or so-called instantons.

Let us consider the following action
\begin{eqnarray}
S = \int dx^{4} \sqrt{-g} \left[ \frac{1}{16\pi} \mathcal{R} - \frac{1}{2} \left( \nabla \phi \right)^{2} \right] + \int_{\partial \mathcal{M}} \frac{\mathcal{K} - \mathcal{K}_{o}}{8\pi} \sqrt{-h} dx^{3},
\end{eqnarray}
where $\mathcal{R}$ is the Ricci scalar, $\mathcal{K}$ is the Gibbons-Hawking boundary term, and $\mathcal{K}_{o}$ is the Gibbons-Hawking boundary term for the periodically identified Minkowski \cite{Gibbons:1976ue}. Note that $\phi$ is a free scalar field and hence there is no contribution of the scalar field to the on-shell action. For any field combinations, the on-shell Euclidean action becomes
\begin{eqnarray}\label{eq:prob}
S_{\mathrm{E}} = - \int_{\partial \mathcal{M}} \frac{\mathcal{K} - \mathcal{K}_{o}}{8\pi} \sqrt{+h} dx^{3} + \left( \mathrm{contribution\; at\; horizon} \right).
\end{eqnarray}
The probability of an instanton process is $P \sim e^{-2 B}$, where
\begin{eqnarray}
B = S_{\mathrm{E}}(\mathrm{solution}) - S_{\mathrm{E}}(\mathrm{background}).
\end{eqnarray}

We impose the classicality condition, i.e., the reality condition of the scalar field at the future infinity. Then the solution must be complex-valued in the bulk region, but we do not physical observe them. The complex-valued scalar field does not contribute to the on-shell action, but due to the energy conservation, there appears a cusp at the horizon of the Euclidean geometry \cite{Chen:2018aij}. After the regularization of the cusp, we obtain the correct on-shell action \cite{Gregory:2013hja}
\begin{eqnarray}
2B = \frac{\mathcal{A}}{4} - \frac{\mathcal{A}'}{4} = 4\pi \left( M^{2} -M'^{2} \right),
\end{eqnarray}
where $\mathcal{A}$ and $M$ are the areal radius and mass of the initial black hole, respectively, while the primed ($'$) quantities denote for the final black hole.

In the end, if $M' = M - \omega$ with a very small $\omega \ll M$, then
\begin{eqnarray}
2B = 8\pi M \omega
\end{eqnarray}
which is the consistent result of Hawking temperature. On the other hand, there exist a wide spectrum of instantons, e.g., $M' = 0$ \cite{Chen:2018aij}. In this case, the probability is exponentially suppressed and no more dominated, but the resulting geometry is trivial, e.g., there exists no horizon nor singularity \cite{Sasaki:2014spa}. Therefore, we can reasonably conclude that the existence of a tunneling channel toward a trivial geometry is very evident, although the price is the exponentially suppressed low probability.

\section{Revisit the Page curve}

As we have checked the existence of the tunneling channel toward a trivial geometry, i.e., the multi-history condition, we need to demonstrate the late-time dominance condition. In order to do this, we first observe the possible variety of the Euclidean time \cite{Chen:2021jzx}.

Basically, if the shell is static, the contribution from the solution is two parts, where from the bulk integration, we obtain $4\pi M_{+}^{2}$ and from the boundary term at infinity, we obtain one more additional $4\pi M_{+}^{2}$. The last term must be canceled from the boundary term at infinity of the background solution. Eventually, we obtain the result that such a transition is exponentially suppressed and the exponential factor is the same as the entropy of the original black hole. Conceptually, one can write as follows:
\begin{eqnarray}
2B = (\mathrm{bulk\; term\; of\; solution}) + (\mathrm{boundary\; term\; of\; solution}) - (\mathrm{boundary\; term\; of\; background}).
\end{eqnarray}

The Euclidean time of the background cannot be chosen freely. If we started from a Schwarzschild black hole, the Euclidean time period of the background geometry is fixed. However, the Euclidean period of the solution part is different. In principle, if the final boundary is the same, we can consider periodically identified Euclidean time period arbitrarily. For this case, we obtain the following factor:
\begin{eqnarray}
2B = n((\mathrm{bulk\; term\; of\; solution}) + (\mathrm{boundary\; term\; of\; solution})) - (\mathrm{boundary\; term\; of\; background}),
\end{eqnarray}
where $n \geq 1$.

If we assume $n$ is an integer, then we obtain the result of the tunneling probability:
\begin{eqnarray}
\sum_{n=1}^{\infty} e^{-S(2n-1)} = \frac{1}{e^{S} - e^{-S}},
\end{eqnarray}
where $S = 4\pi M_{+}^{2}$.

Therefore, if we only consider two histories, where one is a semi-classical black hole and the other is a trivial geometry without a horizon, the probabilities of each history, $p_{1}$ and $p_{2}$, respectively, becomes as follows:
\begin{eqnarray}
p_{1} &=& \frac{e^{S} - e^{-S}}{1 + e^{S} - e^{-S}},\\
p_{2} &=& \frac{1}{1 + e^{S} - e^{-S}}.
\end{eqnarray}

\begin{figure}
\begin{center}
\includegraphics[scale=0.8]{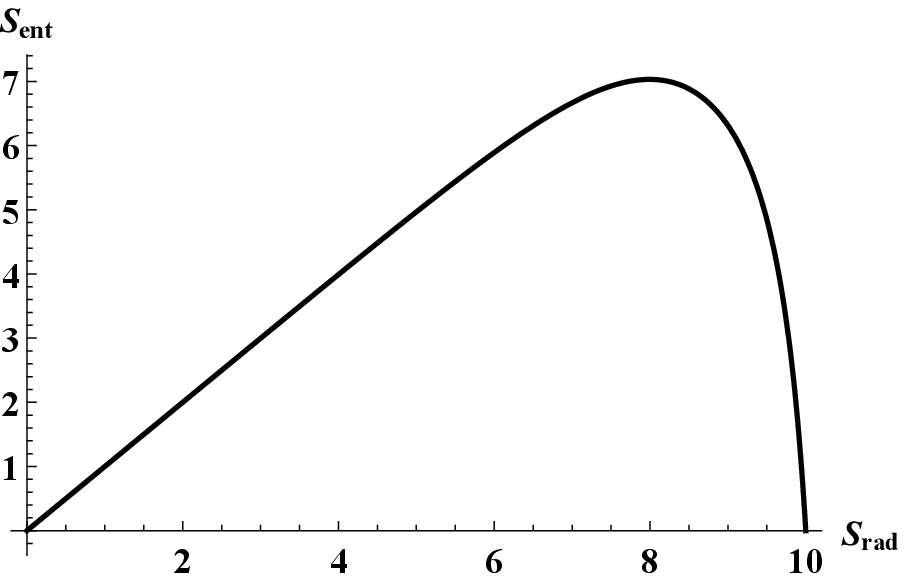}
\includegraphics[scale=0.8]{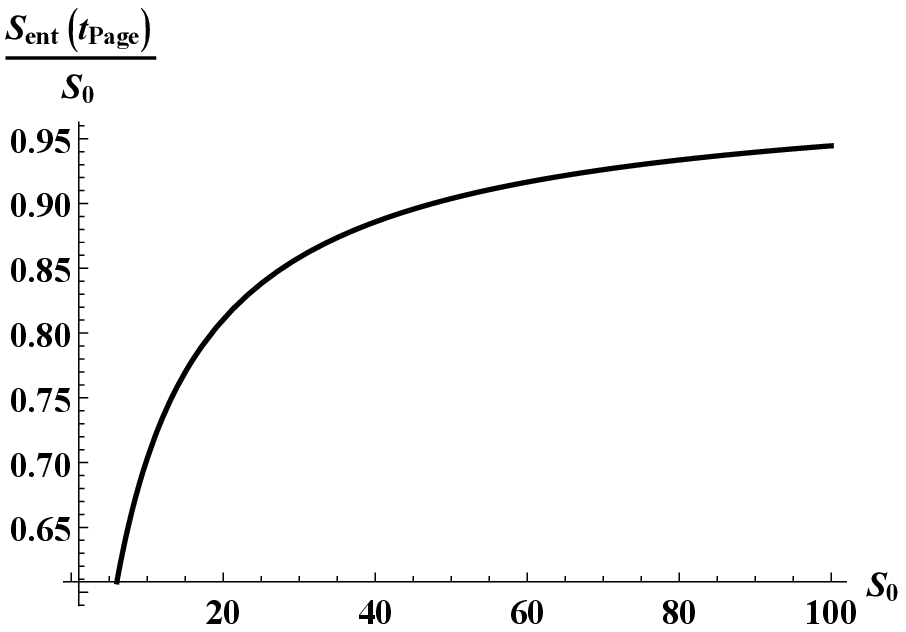}
\caption{\label{fig:ent}Left: An example of the Page curve. The horizontal axis is the entropy of radiation $S_{\mathrm{rad}}\equiv S_{0} - S$ and the vertical axis is the entanglement entropy $S_{\mathrm{ent}}$. Right: At the Page time, we estimate the ratio between the entanglement entropy and the initial entropy. As the initial black hole size increases, the portion of the compressed information increases. (See more details in \cite{Chen:2021jzx}.)}
\end{center}
\end{figure}

We can finally obtain the entanglement entropy as a function of $S$:
\begin{eqnarray}
S_{\mathrm{ent}} &=& \left( S_{0} - S \right) \times p_{1} + 0 \times p_{2} \\
&=& \left( S_{0} - S \right) \left(\frac{e^{S} - e^{-S}}{1 + e^{S} - e^{-S}}\right),
\end{eqnarray}
where $S_{0}$ is the initial entropy of the black hole. Here, for simplicity we assume that the entanglement entropy of $h_{1}$ monotonically increases, while the entanglement entropy of $h_{2}$ goes to zero (because there is no black hole). This explains the unitary evolution of a black hole (see Fig.~\ref{fig:ent}, \cite{Chen:2021jzx}). 

\section{More on the late-time dominance condition}

In the previous section, we only compared with two histories. However, in general, there are infinitely many tunneling channels; also, the number of histories for non-trivial geometries will be much more than those of the trivial geometries. In this regards, can we still sure that the non-singular geometries will eventually be dominated than the geometries with singularities?

Perhaps, the quantum resolution of the singularity, i.e., the quantum gravitational treatment of the singularity will help. The probability of an arbitrary 3-hypersurface will be approximately a multiplication between the probability of the inside and the outside the black hole \cite{Bouhmadi-Lopez:2019kkt}. Of course, in general, the probability will be very complicated, but if we must provide the DeWitt boundary condition for the singularity \cite{DeWitt:1967yk}, the probability for inside the horizon must approach zero; hence, the entire 3-hypersurface with a singularity will approach zero, too.

Interestingly, there are some evidences that the DeWitt boundary condition for the black hole singularity is very generic.
\begin{itemize}
\item[--1.] The quantum gravitational investigation for inside the horizon shows that there exists a quantum bouncing surface \cite{Bouhmadi-Lopez:2019kkt}. Therefore, at this surface, we can assign the DeWitt boundary condition; and one can interpret that the annihilation-to-nothing process happens around the surface.
\item[--2.] Due to the time-symmetry of the spacetime, in various loop quantum gravity inspired black hole models, there exists bouncing points near the spacelike singularity or spacelike hypersurface \cite{Bodendorfer:2019xbp}, not only static black holes, but also gravitational collapsing cases \cite{Brahma:2021xjy}. One can interpret the quantum bouncing point as the hypersurface for the DeWitt boundary condition.
\item[--3.] In addition to them, due to the BKL conjecture of the singularity, there exists an argument that the vanishing boundary condition of the Wheeler-DeWitt equation is necessary \cite{Perry:2021mch}.
\end{itemize}

Therefore, by integrating these arguments, we conclude that the late-time dominance condition is quite reasonable and general, which is justified by conservative quantum gravitational approaches.

\section{Future perspectives}

In this paper, we summarized several ideas. First, the unitary Page curve can be explained, if two conditions (the multi-history condition and the late-time dominance condition) are satisfied. Second, in the Euclidean path integral approach, the multi-history condition is quite evident. Third, the late-time dominance condition can be supported by considering the multiple Euclidean time period; also, the late-time dominance can be further supported by the DeWitt boundary condition inside the black hole.

If all ideas are working consistently, we can understand the information loss paradox in a deeper way. It is a very interesting question that what is the relation to the island conjecture and the replica wormholes. We believe that the island conjecture is deeply related to the previous two conditions. However, it is less clear what is the corresponding geometry in terms of the state-level (Euclidean) path integral. At this point, we need to investigate further between these two approaches.

Also, it is worthwhile to find a way to (theoretically or experimentally) confirm ideas, whether the replica wormholes exist or the other Euclidean tunneling channels do the same role instead of the replica wormholes, etc. We leave these interesting topics for the future research projects.

\begin{acknowledgments}
The author would like to thank for stimulated discussions with Suddhasattwa Brahma, Che-Yu Chen, Pisin Chen, Misao Sasaki, and Junggi Yoon. This work is supported by the National Research Foundation of Korea (Grant no.: 2021R1C1C1008622, 2021R1A4A5031460).
\end{acknowledgments}

\newpage

\nocite{*}

\bibliographystyle{aipprocl}

\end{document}